# Rutile $GeO_2$ Thin Films Grown by Pulsed Laser Deposition on Differently Oriented Sapphire Substrates for Solar-Blind UV-C Photodetection

*Eriks Dipans[1*], Edvards Strods[1], Sven Oras[2], Annamarija Trausa[1], Katrina Laganovska[1], Edgars Vanags[1], Jevgenijs Gabrusenoks[1], Boris Polyakov[1], Esme Isik[1,3], Edgars Butanovs[1*]*

[*]eriks.dipans@cfi.lu.lv; edgars.butanovs@cfi.lu.lv

[1]Institute of Solid State Physics, University of Latvia, Kengaraga street 8, Riga, Latvia

[2]Institute of Technology, University of Tartu, Nooruse 1, Tartu, 50411, Estonia

[3]Department of Electronics and Automation, Malatya Turgut Ozal University, 44700, Malatya, Türkiye

## Abstract

Rutile germanium oxide (r-$GeO_2$) is an emerging ultrawide-bandgap semiconductor with promising properties for solar-blind ultraviolet photodetection, yet the heteroepitaxial growth and photoconductive properties of phase-pure r-$GeO_2$ thin films remain largely unexplored. Here, we systematically investigate the influence of sapphire substrate orientation on the pulsed laser deposition growth and UV-C photodetection performance of r-$GeO_2$ thin films. Phase-pure r-$GeO_2$ was stabilized on r-, a-, and m-plane sapphire under equivalent growth conditions, yielding preferential (101) orientation on r- and a-plane sapphire and (002) orientation on m-plane sapphire. The (002)-oriented film exhibited the highest crystalline alignment, with a rocking-curve full width at half-maximum of 0.46°, together with an improved film–substrate interface and reduced surface roughness. Metal–semiconductor–metal photodetectors fabricated from all three films exhibited solar-blind UV-C photoresponse, with the (002)-oriented device reaching a peak responsivity of 1.60 A/W at 220 nm and a solar-blind rejection ratio ($R_{peak}/R_{280}$) of 34.1 at 5 V bias. The enhanced responsivity was accompanied by a slower temporal response, suggesting a contribution from defect-mediated photoconductive gain. An optical absorption edge of ~5.1 eV and a correspondingly blue-shifted photoconductive onset relative to the fundamental bandgap of ~4.7–4.8 eV were observed for the r-$GeO_2$ films. This suppression of photogeneration near the fundamental band edge effectively extends the intrinsic spectral selectivity deeper into the UV-C region, highlighting r-$GeO_2$ as a promising material for next-generation solar-blind photodetectors.

## Introduction

As the demand for ultrawide-bandgap (UWBG) semiconductors increases, there is growing interest in emerging materials beyond established UWBG systems such as $Ga_2O_3$, AlGaN, and AlN. UWBG semiconductors offer high breakdown fields and intrinsic selectivity to deep-ultraviolet radiation, making them attractive for applications ranging from high-power electronics to solar-blind UV-C optoelectronics [1–3]. Rutile-phase germanium oxide (r-$GeO_2$) has recently emerged as a promising UWBG semiconductor due to its bandgap of around 4.7 eV, predicted electron and hole mobilities of around 300 $cm^2$/Vs and 30 $cm^2$/Vs, respectively, and a thermal conductivity of 52.9 W/mK, nearly twice that of $Ga_2O_3$ [4–6]. Importantly, theoretical studies also predict the possibility of ambipolar doping in r-$GeO_2$ [7].

Solar-blind photodetectors are insensitive to solar radiation reaching the Earth's surface (280–3000 nm), enabling selective detection of UV-C radiation (200–280 nm). Potential applications include environmental and industrial monitoring, such as flame and corona-discharge detection, as well as space science, military technologies, and optical communications [3,8,9]. Current commercial solar-blind photodetectors are based either on bulky, high-voltage vacuum tubes or on low-cost Si devices combined with optical filters, which reduce detection efficiency and lifetime while increasing the minimum detectable light intensity [10–12]. Compared with conventional wide-bandgap semiconductors, such as GaN and SiC, UWBG materials can provide intrinsic solar-blindness due to their sufficiently large bandgaps, enabling filter-free, solid-state UV-C photodetectors with lower operating voltages and power consumption than vacuum-tube-based detectors. Among UWBG oxides, r-$GeO_2$ is particularly promising because its predicted ambipolar dopability could enable native p–n homojunction photodiodes, a capability that remains challenging to achieve in β-$Ga_2O_3$ [4,13]. Despite the growing interest in r-$GeO_2$, experimental demonstrations of photodetectors remain limited. Previous studies have reported photodetection at 405 nm for broadband UV photodetector applications [14], while UV-C detection at a single wavelength of 213 nm has been demonstrated using $GeO_2$ thin film containing both rutile and amorphous components [9]. Thus, the photodetection properties and device potential of phase-pure r-$GeO_2$ thin films have yet to be systematically investigated.

The practical development of r-$GeO_2$ devices strongly depends on the ability to stabilize the rutile phase and achieve high-quality crystalline material. Bulk r-$GeO_2$ single crystals, both unintentionally doped and n-type doped, have been demonstrated with lateral dimensions approaching 1 cm [15], while heteroepitaxial thin film growth has been explored on sapphire or rutile $TiO_2$ substrates using metal-organic chemical vapor deposition (MOCVD), mist-CVD, and pulsed laser deposition (PLD) [4]. Rahaman et al. and Cicconi et al. have reported r-$GeO_2$ thin film growth by MOCVD; however, phase stabilization remains challenging, as amorphous, quartz, and rutile $GeO_2$ phases can coexist depending on the growth conditions [16,17]. Similar challenges associated with $GeO_2$ polymorphism have also been reported for thin films grown by mist-CVD [18,19]. The overlapping growth windows of these polymorphs can therefore result in the coexistence of multiple phases, particularly in CVD-based approaches. This is undesirable for device applications because, unlike the chemically stable rutile phase, amorphous and hexagonal $GeO_2$ are water-soluble, potentially compromising long-term device stability. In contrast, PLD has demonstrated greater success in stabilizing phase-pure r-$GeO_2$ films directly on sapphire substrates without the need for complex buffer-layer structures. Deng et al. demonstrated direct PLD growth of (001)-oriented r-$GeO_2$ on m-plane sapphire and subsequently achieved phase-pure (100)-oriented r-$GeO_2$ on c-plane sapphire, with a rocking-curve full width at half-maximum (FWHM) of 24 arc min (0.4°) [20,21]. More recently, Suzuki et al. systematically investigated the effects of substrate temperature and oxygen pressure on the phase stabilization of PLD-grown r-$GeO_2$ on c-plane sapphire [22]. Their results showed that films deposited below 500 °C remained amorphous, while stabilization of the rutile phase required reducing growth conditions with oxygen pressures below 0.1 Pa. The resulting films exhibited sixfold rotationally distributed columnar domains, with residual amorphous $GeO_2$ between the crystalline domains. Despite these advances, PLD growth of r-$GeO_2$ remains relatively unexplored, and the role of sapphire substrate orientation in phase stabilization, crystalline quality, and film morphology remains insufficiently understood.

In this work, we systematically investigate the influence of sapphire substrate orientation on the heteroepitaxial growth and photodetection properties of r-$GeO_2$ thin films. Phase-pure r-$GeO_2$ films are grown by PLD on r-, a-, and m-plane sapphire under equivalent deposition conditions, enabling a direct comparison of their structural, morphological, optical, and photoconductive properties. Solar-blind photodetectors fabricated from each film are further evaluated, establishing direct relationships between substrate orientation, resulting film properties, and UV-C photodetection performance and providing practical guidance for substrate selection in r-$GeO_2$-based devices.

## Materials and Methods

Rutile $GeO_2$ thin films were heteroepitaxially grown on r-, a-, and m-plane sapphire substrates (Biotain Crystal Co., Ltd.) using a PLD system (Twente Solid State Technology) equipped with a nanosecond KrF excimer laser ($\lambda = 248$ nm) and a radiative heater. A c-plane sapphire substrate was also tested but did not yield detectable crystalline r-$GeO_2$ under the comparative growth conditions. A ceramic $GeO_2$ target (AbleTarget Ltd., >65% density) was ablated for 12 000 pulses at a laser fluence of approximately 1.5 J/cm$^2$ and a repetition rate of 8 Hz. The distance between the target and substrate was 54 mm. The substrate temperature was varied between 325 and 425 °C, with the temperature measured at the edge of the sample holder, approximately 1 cm from the substrate. The $Ar/O_2$ background gas composition was varied using flow-rate combinations of 1/4, 3/2, and 5/0 sccm, respectively, while the background pressure was varied between 0.1 and 1 Pa. Among the investigated gas compositions, an $Ar/O_2$ flow ratio of 3:2 at a total pressure of 0.1 Pa provided the most favorable conditions for stabilizing the rutile phase and was therefore used for the comparative growth experiments. The resulting films had thicknesses ranging from 250 to 350 nm, as determined by spectroscopic ellipsometry (RC2–XI, J.A. Woollam).

As-fabricated films were characterized by X-ray diffraction (XRD) using a Rigaku MiniFlex 600 diffractometer (Cu Kα radiation, 40 kV, 15 mA) in Bragg–Brentano θ–2θ geometry for 2θ scans and a PANalytical X'Pert Pro diffractometer (Cu $K\alpha_1$ radiation, 40 kV, 30 mA) for ω (rocking-curve) scans. X-ray photoelectron spectroscopy (XPS) measurements were performed using an ESCALAB Xi spectrometer (Thermo Fisher Scientific). Al Kα radiation (1486 eV) was used as the excitation source. The analyzed area was 650 μm × 100 μm, and the angle between the analyzer and the sample surface was 90°. An electron gun was used to perform charge compensation; however, no sputter-cleaning was performed prior to the measurements. The base pressure during spectral acquisition was better than $10^{-5}$ Pa. Micro-Raman measurements were carried out with a TriVista 777 confocal Raman spectrometer (Spectroscopy & Imaging GmbH; Princeton Instruments monochromator) equipped with a continuous-wave single-frequency laser emitting at 532 nm.

The surface morphology of the films was examined using a scanning electron microscope (SEM; Helios 5 UX, Thermo Fisher Scientific) at an accelerating voltage of 5 kV and a beam current of 13 pA using a through-the-lens detector. Atomic force microscopy (AFM, Veeco Dimension Edge) was performed in tapping mode using a non-contact AFM probe (NCHR, Nanoworld; force constant 42 N/m). Cross-sectional lamellae were prepared using the $Ga^+$ focused ion beam (FIB) system of the Helios 5 UX operated at 30 kV. Prior to FIB milling, the samples were coated with Au, followed by in situ deposition of an approximately 2 μm-thick protective Pt layer over the region of interest using the gas injection system. Scanning transmission electron microscopy (STEM) imaging of the lamellae was subsequently performed using the same microscope at an accelerating voltage of 30 kV and a beam current of 0.1 nA with a STEM 3+ detector operated in bright-field (BF) mode. The atomic structure was further investigated by transmission electron microscopy (TEM, FEI Tecnai GF20) operated at an accelerating voltage of 200 kV.

The optical transmittance and reflectance of the films in the range 200–2500 nm were measured by an Agilent Cary 7000 spectrophotometer. The sample was positioned at an incidence angle of 6° relative to the *P*-polarized beam, and the detector was placed at an angle of 180° behind the sample to measure the transmittance and at 12° in front of the sample to measure the specular reflectance. Photoluminescence (PL) measurements were performed at room temperature using a YAG laser FQSS266 (CryLas GmbH) 4th harmonic at 266 nm (4.66 eV). The PL spectra were recorded using a Horiba iHR320 spectrograph equipped with a CCD camera (Andor DV420A-BU2).

Photodetector devices were fabricated by patterning interdigitated electrodes on the as-grown r-$GeO_2$ films without mesa isolation using conventional photolithography, followed by thermal evaporation of Ti/Au (15/75 nm) contacts. The detailed interdigitated electrode geometry is shown in Fig. S1 and consists of 600 μm-long, 15 μm-wide electrode fingers separated by 14 μm gaps, resulting in a total exposed semiconductor area between the electrodes of approximately $1.008\cdot10^{-7}$ $m^2$, used as the effective illuminated area for responsivity calculations. The electrode pattern was transferred using a SUSS MicroTec

MA/BA6 Gen4 mask aligner in soft-contact mode and a bilayer resist stack consisting of LOR-3A (Micro Resist Technology GmbH) and AZ1518 (Merck Performance Materials GmbH). Photoconductivity and current–voltage (I–V) measurements were performed using a microprobe station connected to a Keithley 2450 source-measure unit (Tektronix). Time-resolved photoresponse measurements were carried out using an SR570 low-current preamplifier (Stanford Research Systems) coupled to a TDS2004B oscilloscope (Tektronix). Optical excitation was provided by an EQ-77X laser-driven light source (LDLS, Energetiq) coupled to an ML44 monochromator (Solar Laser Systems, 9 nm spectral bandwidth) and an SH05 optical beam shutter (Thorlabs). The illumination spot, with an area of approximately $2.843 \cdot 10^{-6}$ $m^2$, was centered on the device and fully covered the interdigitated electrode region. All measurements were carried out in air at room temperature.

## Results and Discussion

For r-$GeO_2$ films grown at 375 °C using an Ar:$O_2$ flow ratio of 3:2, XRD θ–2θ scans revealed a strong dependence of the preferential growth orientation on the sapphire substrate plane (Fig. 1). Films grown on r- and a-plane sapphire exhibited the r-$GeO_2$ (101) Bragg reflection at $2\theta \approx 37.5°$ and the higher-order (202), whereas the film grown on m-plane sapphire exhibited the (002) reflection at 2θ ≈ 65.2°. Both reflections are consistent with the rutile $GeO_2$ reference pattern (ICDD-PDF 00-035-0729). Apart from reflections originating from the sapphire substrates (corundum $Al_2O_3$, ICDD-PDF 00-046-1212), no additional diffraction peaks were observed, confirming the formation of phase-pure r-$GeO_2$ films on all three sapphire orientations. XRD patterns of the corresponding bare sapphire substrates are shown in Fig. S2. It is worth noting that the film grown on c-plane sapphire exhibited no diffraction peaks other than those of the substrate, indicating the absence of detectable crystalline $GeO_2$ phases; this sample was therefore excluded from further investigation. In contrast, the atomic arrangement of Ge sites in r-$GeO_2$ shows favorable structural matching with oxygen sites exposed on the a- and m-plane sapphire surfaces, resulting in relatively low lattice mismatch and facilitating heteroepitaxial growth [4,23]. The lack of crystalline $GeO_2$ on c-plane sapphire under the present growth conditions may therefore be associated with a less favorable epitaxial relationship between tetragonal r-$GeO_2$ and the hexagonally terminated c-plane $Al_2O_3$ surface. Furthermore, rocking-curve (ω-scan) measurements yielded FWHM values of 1.20°, 0.61°, and 0.46° for the films grown on r-, a-, and m-plane sapphire, respectively (insets of Fig. 1(a–c)). The lowest FWHM obtained for the (002)-oriented film on m-plane sapphire indicates the lowest mosaic spread and highest degree of out-of-plane crystalline alignment among the investigated films. The substrate temperature was found to strongly influence the crystalline alignment, with 375 °C yielding the minimum rocking-curve FWHM of 0.46° for r-$GeO_2$ films grown on m-plane sapphire (see Fig. S3). Growth at both lower and higher temperatures resulted in broader rocking curves, indicating increased mosaic spread and establishing 375 °C as the optimal growth temperature within the investigated range. Notably, the optimal growth temperature of 375 °C is lower than the 500–600 °C reported in previous PLD studies of r-$GeO_2$ [21,22], which may partly reflect differences in substrate temperature calibration, target–substrate distance, and/or other system-specific deposition conditions affecting the energy and flux of the arriving ablated species.

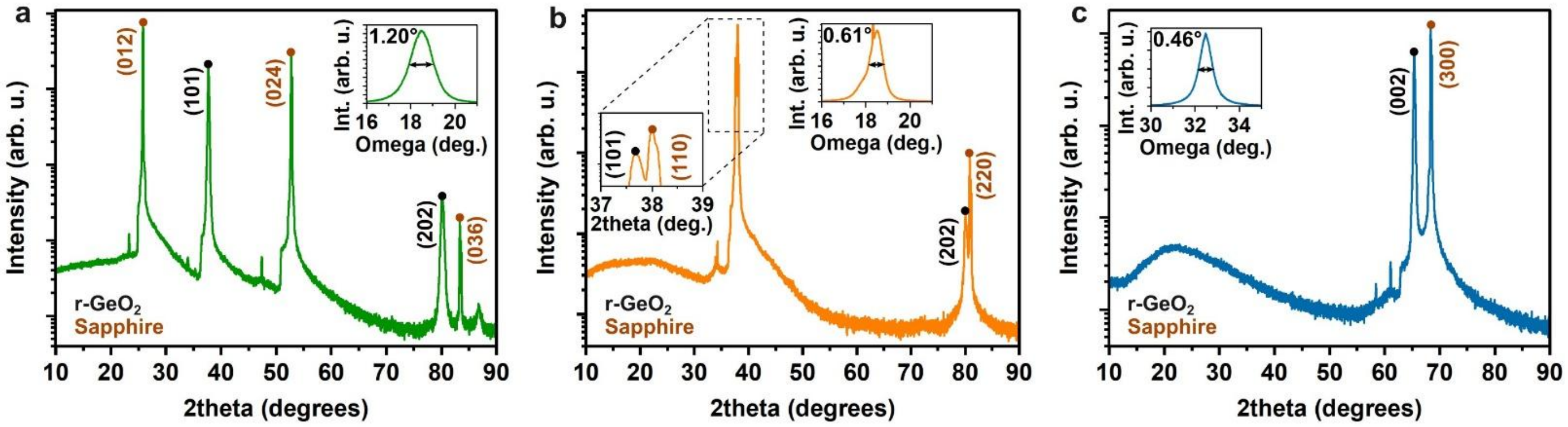

**Figure 1.** XRD θ–2θ scans of r-$GeO_2$ thin films grown on (a) r-plane, (b) a-plane, and (c) m-plane sapphire substrates. Insets show the corresponding rocking curves and FWHM values of the r-$GeO_2$ (101), (101), and (002) reflections, respectively.

Cross-sectional BF-STEM images (Fig. 2(a,e,i)) of the films grown on r-, a-, and m-plane sapphire, respectively, reveal a columnar microstructure in all r-$GeO_2$ films, with the columnar grains gradually widening toward the film surface. Notably, the columns in the film grown on r-plane sapphire are inclined relative to the substrate surface normal (Fig. 2(a)), whereas those grown on a- and m-plane sapphire are predominantly oriented perpendicular to the substrate surface. Furthermore, the film thicknesses determined by spectroscopic ellipsometry were 268, 267, and 338 nm for the films grown on r-, a-, and m-plane sapphire, respectively, in good agreement with the film thicknesses observed in the cross-sectional STEM images. High-resolution (HR) TEM images of the film–substrate interfaces (Fig. 2(b,f,j)) reveal pronounced differences in the initial growth of r-$GeO_2$ on the three sapphire orientations. On r-plane sapphire (Fig. 2(b)), r-$GeO_2$ nucleates with considerable structural disorder, with distorted lattice regions and locally varying crystallographic orientations already present close to the interface. This imperfect initial alignment is consistent with the larger mosaic spread observed by rocking-curve measurements and likely contributes to the inclined columnar growth observed throughout the film. The interface with a-plane sapphire (Fig. 2(f)) exhibits improved structural ordering, although localized lattice distortions and extended defects are still visible near the interface. In contrast, r-$GeO_2$ grown on m-plane sapphire (Fig. 2(j)) exhibits a substantially more ordered film–substrate interface, with comparatively few visible structural defects. These observations are consistent with the progressive decrease in rocking-curve FWHM from r- to a- and m-plane sapphire, supporting improved epitaxial alignment of r-$GeO_2$ on m-plane sapphire. HR-TEM images acquired within the columnar grains in the central region of the film (Fig. 2(c,g,k)) reveal well-resolved lattice fringes and improved lattice ordering compared with the film–substrate interfaces. For the films grown on

r- and a-plane sapphire (Fig. 2(c,g)), a [101] growth direction is identified, with an interplanar spacing of approximately 2.5 Å measured along this direction, consistent with the reference crystallographic data for r-$GeO_2$ (ICDD-PDF 00-035-0729). In contrast, the film grown on m-plane sapphire (Fig. 2(k)) exhibits a [001] growth direction, while the (101) lattice planes with a spacing of approximately 2.5 Å are also clearly resolved. The crystallographic orientations are further supported by the corresponding selected-area electron diffraction (SAED) patterns (Fig. 2(d,h,l)) and are consistent with the (101) and (002) preferential orientations identified by XRD. A detailed crystallographic analysis, including the alignment of the r-$GeO_2$ crystal structure with the HR-TEM images and indexing of the SAED patterns acquired from the film, film–substrate interface, and sapphire substrate, is provided in Figure S4.

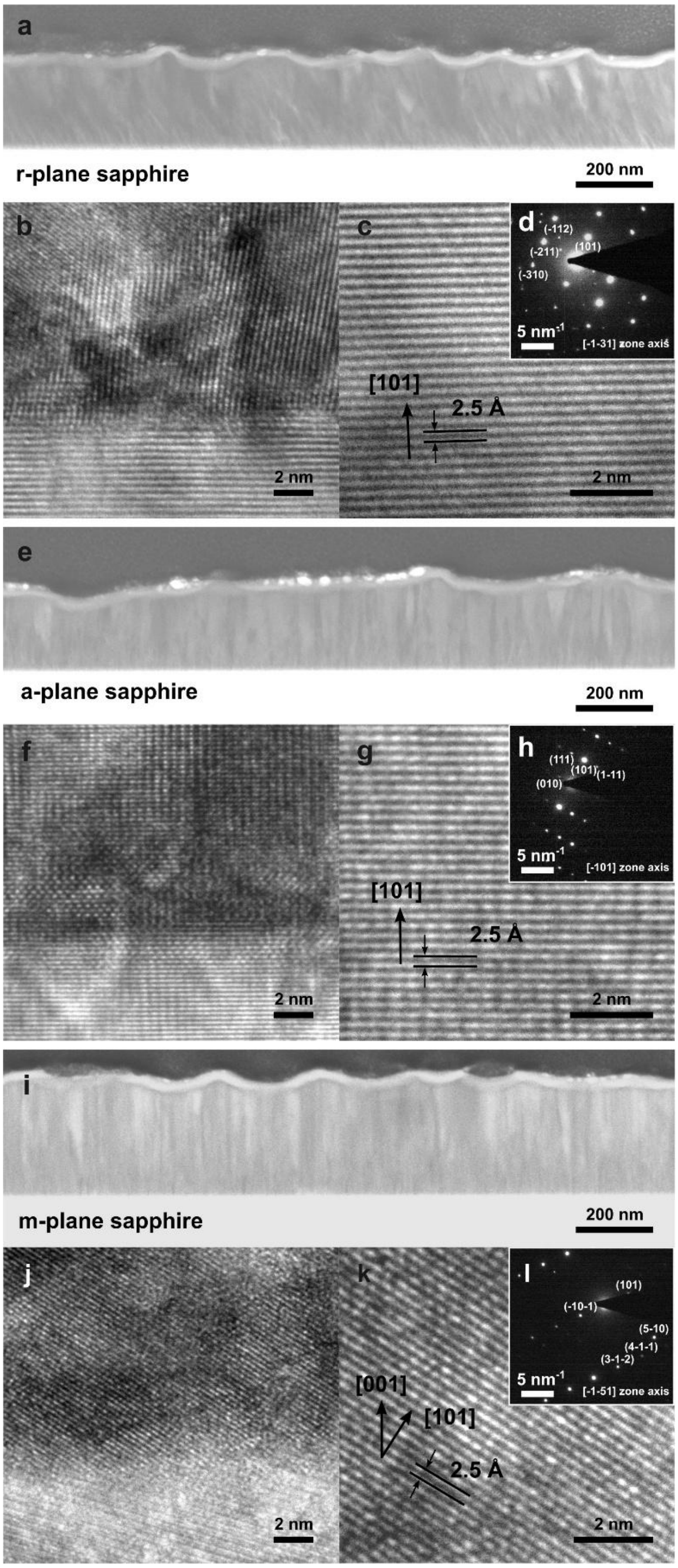

**Figure 2.** Cross-sectional STEM images, HR-TEM images of the film–substrate interface and central region of the film, and corresponding SAED patterns of r-$GeO_2$ thin films grown on (a–d) r-plane, (e–h) a-plane, and (i–l) m-plane sapphire substrates, respectively.

The surface morphology of the r-$GeO_2$ films was further investigated by SEM and AFM, as shown in Fig. 3. Pronounced differences in surface morphology are observed depending on the sapphire orientation. The films grown on r- and a-plane sapphire exhibit relatively rough surfaces, with root-mean-square (RMS) roughness values of 6.90 and 6.15 nm, respectively, whereas the film grown on m-plane sapphire shows a substantially smoother surface with an RMS roughness of 2.70 nm. The m-plane sample also exhibits smaller and more uniformly distributed surface features compared with the coarser morphology observed for the r- and a-plane samples. This trend follows the decrease in rocking-curve FWHM from 1.20° for r-plane to 0.61° for a-plane and 0.46° for m-plane sapphire, indicating that improved crystalline alignment is accompanied by a smoother and more uniform surface morphology. The AFM image of the film grown on r-plane sapphire additionally reveals elongated, approximately parallel surface features that are not observed for the other substrate orientations. Their presence may be related to the inclined columnar microstructure observed by cross-sectional STEM (Fig. 2(a)), since such inclined columns can produce an anisotropic surface projection/coalescence morphology; however, their precise crystallographic origin cannot be determined from the present measurements. Films grown outside the conditions optimized for crystalline quality generally exhibited higher surface roughness (Fig. S3), which was also considered during the optimization of the deposition conditions.

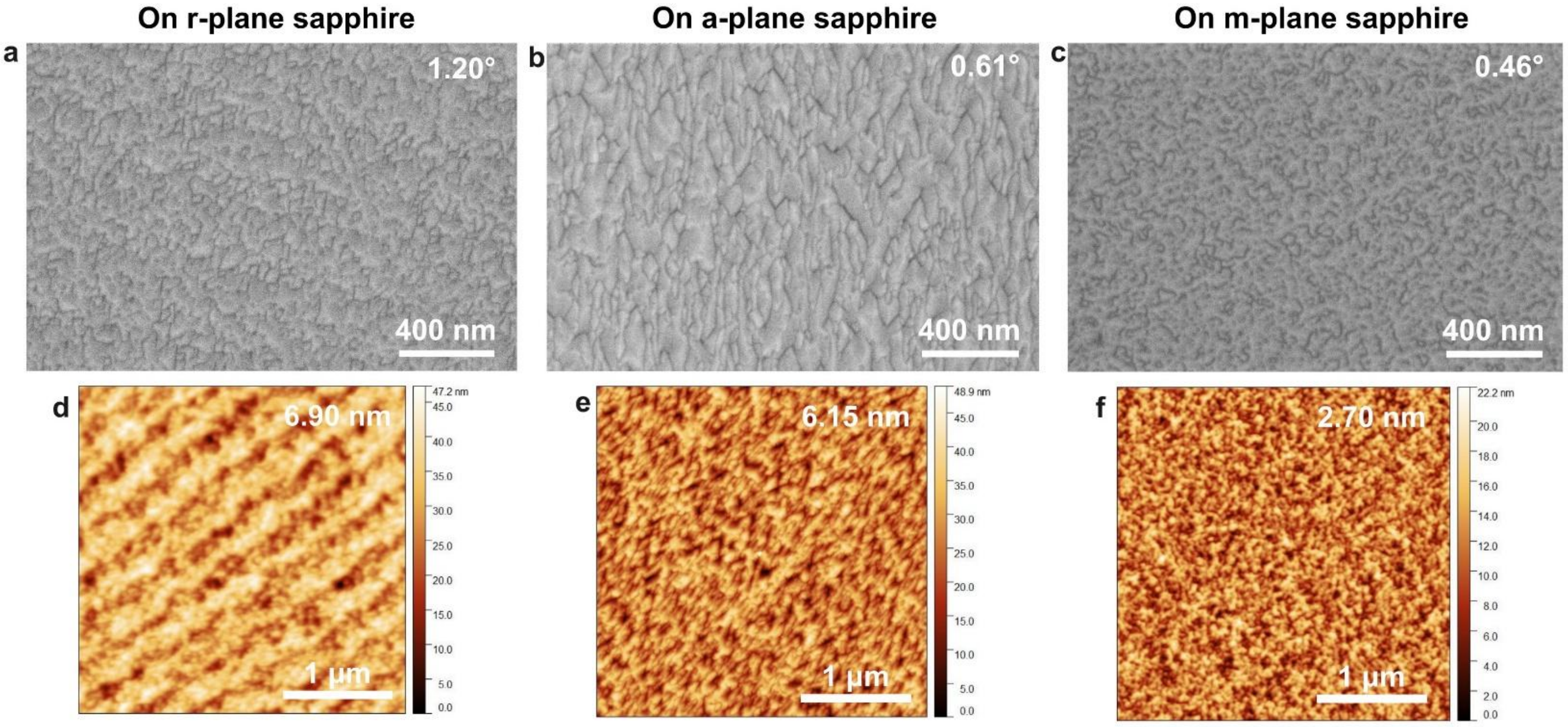

**Figure 3.** Surface morphology of r-$GeO_2$ thin films grown on r-, a-, and m-plane sapphire substrates. (a–c) SEM images and (d–f) corresponding AFM topography images, respectively. The rocking-curve FWHM and RMS surface roughness values are indicated for each film.

The phase composition of the r-$GeO_2$ films was further examined by micro-Raman spectroscopy. The Raman spectra of all three films (Fig. 4(a,e,i)) exhibit the characteristic $B_{1g}$ and $A_{1g}$ Raman modes of rutile $GeO_2$ at approximately 168.6 and 698.5 $cm^{-1}$, respectively [17,23], further confirming stabilization of the rutile phase. No amorphous $GeO_2$ phase was detected [17], while the additional peaks originate from the sapphire substrates and are marked accordingly. Notably, the $A_{1g}$ mode of the film grown on m-plane sapphire is red-shifted by approximately 2 $cm^{-1}$ to 696.6 $cm^{-1}$, whereas no measurable shift of the $B_{1g}$ mode is observed. The selective shift of the $A_{1g}$ mode is consistent with anisotropic residual strain and associated distortion of the $GeO_6$ octahedra in the (002)-oriented film. In structurally analogous rutile oxides such as $TiO_2$, the $A_{1g}$ vibration is particularly sensitive to changes in the metal–oxygen bonding within the rutile octahedra, and its softening has been associated with anisotropic lattice strain [24,25]. The observed $A_{1g}$ redshift, together with the unchanged $B_{1g}$ position, may therefore reflect orientation-dependent lattice distortion in the r-$GeO_2$ film on m-plane sapphire. However, the relatively small magnitude of the shift does not allow an unambiguous strain assignment based on Raman spectroscopy alone.

The surface chemical composition and bonding states of the r-$GeO_2$ films were investigated by XPS. Survey spectra of all three samples (Fig. 4(b,f,j)) show Ge and O as the dominant elements, together with adventitious carbon originating from surface contamination, while no additional elements were detected. Quantitative XPS analysis yielded an approximate surface Ge:O ratio of 1:1.7 for all three substrate orientations, indicating a slightly oxygen-deficient surface composition relative to nominal $GeO_2$ stoichiometry. However, due to the surface-sensitive nature of XPS, this ratio cannot be directly interpreted as a quantitative bulk oxygen-vacancy concentration. High-resolution Ge 3d and O 1s core-level spectra, shown in Fig. 4(c,d,g,h,k,l), were calibrated to the adventitious C 1s peak at 284.8 eV. The Ge 3d spectra exhibit the characteristic $Ge^{4+}$ contribution at approximately 32.3 eV for Ge $3d_{5/2}$, with a spin–orbit splitting of ~0.6 eV between the Ge $3d_{5/2}$ and Ge $3d_{3/2}$ components. The O 1s spectral peaks are centered at approximately 531.1 eV and are assigned predominantly to lattice oxygen in Ge–O bonding. These binding energies are consistent with previously reported values for r-$GeO_2$ thin films [9,20], and indicate similar surface chemical states for the films grown on the different substrate orientations. Normalized overlays of the Ge 3d and O 1s XPS spectra are presented in Fig. S5. While the Ge 3d peaks exhibit negligible differences in binding energy, the O 1s peaks show a more pronounced substrate-dependent shift, particularly for the m-plane sample, consistent with variations in the near-surface oxygen chemical environment and potentially oxygen-related defect states [26,27].

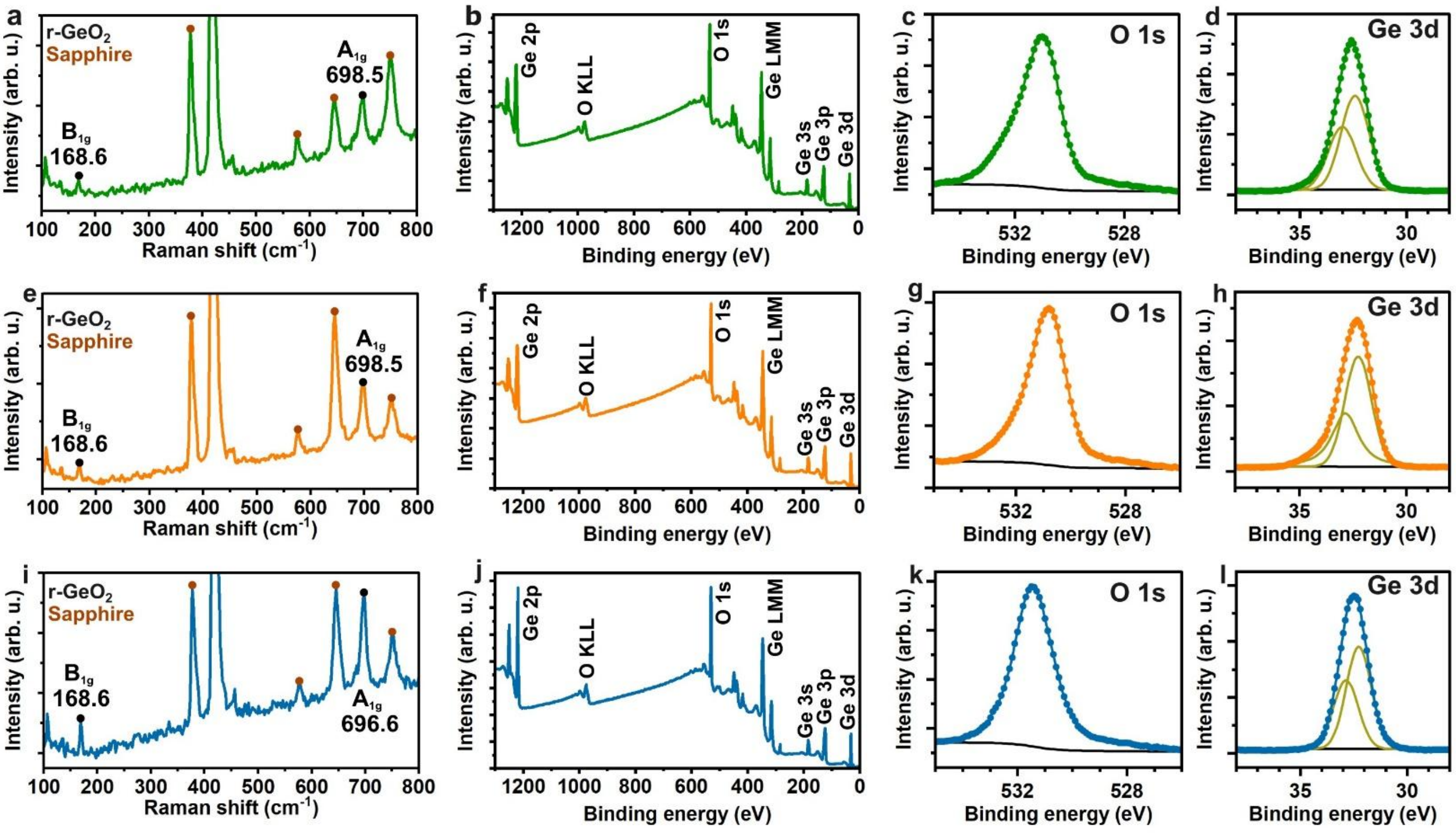

**Figure 4.** Micro-Raman and XPS characterization of r-$GeO_2$ thin films grown on differently oriented sapphire substrates. Raman spectra, XPS survey spectra, high-resolution O 1s spectra, and high-resolution Ge 3d spectra are shown for films grown on (a–d) r-plane, (e–h) a-plane, and (i–l) m-plane sapphire, respectively.

The optical properties of the r-$GeO_2$ films were investigated by UV–Vis-NIR spectrophotometry, with the corresponding absorptance, transmittance, and reflectance spectra provided in Fig. S6. Tauc analysis (Fig. 5(a)) yielded apparent optical bandgap values of approximately 5.13 eV for the films grown on r- and m-plane sapphire and 5.15 eV for the film on a-plane sapphire. These values are noticeably higher than the theoretically predicted fundamental bandgap of r-$GeO_2$ of approximately 4.7–4.8 eV [13]. Similar discrepancies between the fundamental electronic bandgap and experimentally observed optical absorption edge have been reported for both r-$GeO_2$ thin films and bulk single crystals [22,28]. For PLD-grown r-$GeO_2$ on c-plane sapphire, Deng et al. reported an optical bandgap of 5.12 eV and proposed lattice strain as a possible origin of the increased value [20]. Although the small $A_{1g}$ Raman shift observed for the (002)-oriented film in the present study is consistent with some residual lattice strain, no appreciable shift of the r-$GeO_2$ XRD reflections was observed.

Therefore, strain alone appears unlikely to account for the ~0.3–0.4 eV difference between the predicted fundamental bandgap and the measured optical absorption edge. Theoretical calculations by Mengle et al. provide an alternative explanation, showing that the fundamental electronic transition near 4.7–4.8 eV is optically weak or dipole-forbidden, while the first allowed transition for E ⊥ c occurs at 5.04 eV, resulting in substantial absorption only above ~5 eV [13]. This suppression of near-band-edge absorption shifts the effective absorption onset deeper into the UV-C region, which is advantageous for solar-blind photodetection. Consistent with this prediction, Galazka et al. reported strongly anisotropic absorption in bulk r-$GeO_2$, with onsets of approximately 5.0 eV for E ⊥ c and 5.5 eV for E ∥ c [29]. Although *P*-polarized light was used in the present study, the measurements were not configured to separately resolve the E ⊥ c and E ∥ c optical responses, potentially explaining the similar apparent optical bandgaps of ~5.1 eV despite the different (101) and (002) preferential film orientations. Overall, the measured ~5.1 eV absorption edge likely corresponds to the onset of strong allowed optical transitions rather than the fundamental electronic bandgap of r-$GeO_2$, although further dedicated polarization-resolved measurements with respect to the crystallographic c-axis are required to fully clarify its anisotropic optical absorption behavior. A similar distinction between the fundamental electronic gap and the onset of strong optical absorption is known in other oxide semiconductors; for example, bixbyite $In_2O_3$ exhibits a fundamental gap of ~2.7–2.9 eV but a strong optical absorption edge near ~3.7–3.8 eV because of the weak or dipole-forbidden character of the band-edge transition [30,31], while rutile $TiO_2$ likewise exhibits weak near-edge transitions associated with optical selection rules [32].

PL spectra of the r-$GeO_2$ films measured at room temperature are shown in Fig. 5(b). All three films exhibit broad blue–green emission substantially below the fundamental bandgap, indicating that the observed luminescence is dominated by defect-related recombination rather than near-band-edge emission. Similar behavior has recently been observed in bulk r-$GeO_2$ single crystals, where intense blue and green PL bands centered at approximately 2.95 and 2.38 eV, respectively, were reported in the absence of near-band-edge emission [29]. Previous studies of rutile and rutile-like $GeO_2$ have likewise reported broad emission bands around ~3.0 and ~2.3–2.7 eV associated with oxygen-deficient luminescence centers [33,34]. However, detailed microscopic assignments of individual defect-related emission bands remain limited for r-$GeO_2$, with most established defect assignments originating from studies of amorphous or quartz-like $GeO_2$. The film grown on m-plane sapphire exhibits the most structured spectrum, with components centered at approximately 2.90, 3.15, and 3.30 eV. The bands in the ~2.9–3.2 eV range are therefore tentatively assigned to oxygen-deficiency-related blue/violet emission, while the higher-energy 3.30 eV component likely represents an overlapping defect-related transition rather than direct near-band-edge recombination. The film grown on a-plane sapphire shows a dominant emission near 2.97 eV, consistent with the same broad oxygen-deficiency-related band. The lower-energy ~2.60 eV component indicates a deeper defect-related recombination pathway; calculations for $GeO_2$ have associated emission in this energy range with more complex oxygen-deficient configurations [35], although a unique microscopic assignment cannot be made for the present r-$GeO_2$ films. The most pronounced difference between the investigated films is observed for the m-plane sample, which exhibits a dominant emission component centered at approximately 3.15 eV, accompanied by a weaker shoulder at ~2.90 eV and a higher-energy component at ~3.30 eV. In contrast, the ~3.15 eV emission is substantially weaker in the spectra of the r- and a-plane samples, suggesting differences in the relative populations of optically active defect centers between the films, particularly those associated with oxygen-deficiency-related luminescence. Notably, the integrated PL intensity of the m-plane film is approximately 26 times higher than that of the films grown on r- and a-plane sapphire. Considering that this sample also exhibits the lowest rocking-curve FWHM, the substantially enhanced PL intensity may reflect more efficient radiative recombination and reduced nonradiative losses rather than simply a higher overall defect concentration. It should be noted that rocking-curve FWHM primarily reflects crystalline alignment and mosaicity and is not a direct measure of point-defect concentration.

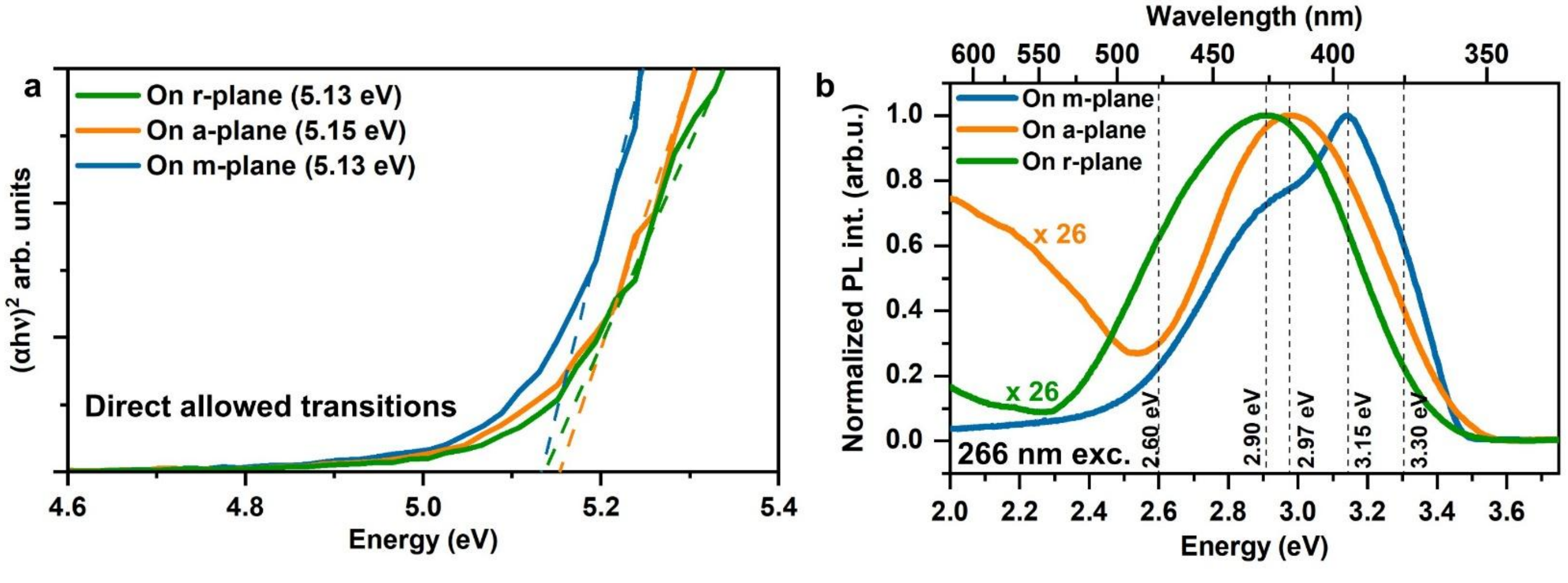

**Figure 5.** Optical properties of r-$GeO_2$ thin films grown on r-, a-, and m-plane sapphire substrates. (a) Tauc plots used to determine the apparent optical bandgap and (b) room-temperature PL spectra measured using an excitation wavelength of 266 nm. The integrated PL intensities of the films on r- and a-plane sapphire are approximately 26 times lower than that of the film on m-plane sapphire.

The photoconductive properties of the fabricated r-$GeO_2$ metal-semiconductor-metal (MSM) photodetectors were evaluated, with the spectral responsivity curves shown in Fig. 6(a–c) and the quantitative performance parameters summarized in Table 1. The spectral responsivity was calculated as $R_\lambda=(I_{on}-I_{dark})/(P_\lambda S)$, where $I_{on}$ and $I_{dark}$ are the currents under illumination and in the dark, respectively, $P_\lambda$ is the incident optical power density at wavelength $\lambda$, and S is the effective illuminated area of the photodetector [36]. The devices fabricated on r-, a-, and m-plane sapphire exhibited peak responsivities of 5.8, 1.8, and 1603.7 mA/W, respectively, at 5 V bias. Most notably, the (002)-oriented r-$GeO_2$ film grown on m-plane sapphire exhibits a peak responsivity nearly three orders of magnitude higher than those of the (101)-oriented films grown on r- and a-plane sapphire. This pronounced difference demonstrates a strong dependence of the photoconductive response on the substrate orientation and the resulting film properties. For all three devices, the spectral response onset is shifted toward shorter wavelengths relative to that expected from the fundamental electronic bandgap of ~4.7–4.8 eV (~260 nm). This behavior is consistent with the optical absorption measurements discussed above and further supports the interpretation that the fundamental transition is optically weak or dipole-forbidden, with efficient photogeneration occurring only at higher photon energies corresponding to the onset of strong allowed absorption. The external quantum efficiency was calculated as $EQE=R_\lambda hc/(e\lambda)$, while the specific detectivity was estimated as $D^*=R_\lambda S^{1/2}/(2eI_{dark})^{1/2}$, where h is Planck's constant, c is the speed of light, and e is the elementary charge. The resulting values for all three devices are summarized in Table 1. The EQE and D* followed the same orientation-dependent trend as the responsivity, reaching 900% and $4.6 \cdot 10^{12}$ Jones, respectively, for the (002)-oriented device on m-plane sapphire. The EQE substantially exceeding 100% provides clear evidence of photoconductive gain, consistent with the pronounced enhancement in responsivity observed for this device. The solar-blind rejection ratio, defined here as $R_\lambda$ at the peak response wavelength divided by $R_\lambda$ at 280 nm, was 9.2, 4.5, and 34.1 for the r-, a-, and m-plane devices, respectively. Similarly, the visible-blind rejection ratio, defined as the peak responsivity divided by $R_\lambda$ at 400 nm, was 41.2, 37.5, and 153.4, respectively. Thus, the m-plane device exhibits substantially improved spectral selectivity in addition to its considerably higher responsivity. Its solar-blind rejection ratio of 34.1 is comparable to values reported for other ultrawide bandgap oxide solar-blind photoconductors [37–40]. Nevertheless, the relatively moderate visible rejection ratios indicate the presence of a sub-bandgap photoresponse, which is consistent with a contribution from defect-related electronic states within the r-$GeO_2$ bandgap.

The insets in Fig. 6(a–c) show the dependence of photocurrent on incident optical power, measured at the peak-responsivity wavelength of each device. The experimental data were fitted using the power-law relation $I_{ph} \sim P^{\gamma}$, where $I_{ph}$ is the photocurrent, P is the incident optical power, and $\gamma$ is the power-law exponent. For an ideal linear photodetector, $\gamma$ approaches unity, whereas sublinear behavior ($\gamma < 1$) is commonly associated with carrier trapping, defect-mediated recombination, and/or saturation of available trap states [41]. The extracted $\gamma$ values were 0.75, 0.79, and 0.71 for the devices fabricated on r-, a-, and m-plane sapphire, respectively, indicating appreciable trap-mediated photoconductive behavior in all three devices.

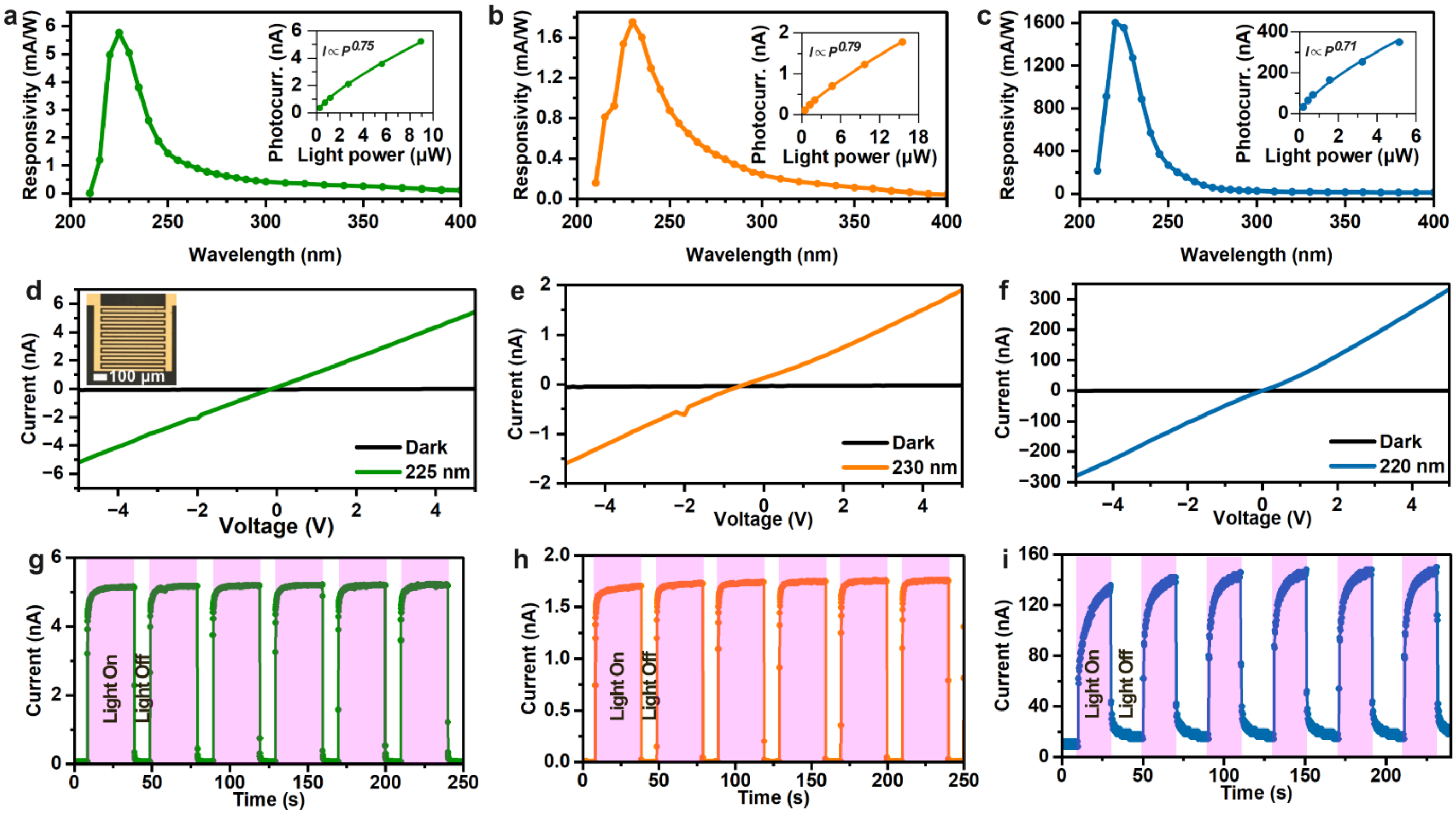

**Figure 6.** Photodetection characteristics of r-$GeO_2$ thin-film MSM devices fabricated on r-, a-, and m-plane sapphire substrates. (a–c) Spectral responsivity, (d–f) dark and UV-illuminated I–V characteristics, and (g–i) time-resolved on–off photoresponse, respectively. Insets in (a–c) show the photocurrent dependence on incident optical power and the corresponding power-law exponent $\gamma$. The inset in (d) shows an optical microscope image of a representative interdigitated MSM device with a 14 μm electrode spacing. The UV illumination wavelengths in (d–f) correspond to the peak responsivity wavelength of each device. Spectral responsivity, power-dependent photocurrent, and time-resolved measurements were performed at a bias voltage of 5 V.

The measured I–V characteristics under dark and UV-illuminated conditions at the respective peak-responsivity wavelengths are shown in Fig. 6(d–f), with the illumination wavelength for each device selected to represent its maximum photoresponse; direct wavelength-dependent comparison among the three devices is provided by the spectral responsivity measurements. All three devices exhibit approximately linear and symmetric I–V characteristics, indicating the absence of strongly rectifying metal/semiconductor interfaces. The dark currents at 5 V were approximately 38, 10, and 750 pA for the devices fabricated on r-, a-, and m-plane sapphire, respectively, indicating substantially higher dark-state conductance of the device based on the (002)-oriented film. A small zero-bias current offset is observed for the device fabricated on a-plane sapphire, which may originate from slight asymmetry between the two metal/semiconductor interfaces or from the measurement setup. The minor deviations from ideal linearity may also be associated with interfacial barriers, as the Ti/Au contacts were deposited directly onto the as-grown r-$GeO_2$ films without subsequent contact annealing.

To determine the temporal response characteristics of the fabricated photodetectors, they were biased at 5 V and periodically illuminated at their respective peak-responsivity wavelengths (Fig. 6(g–i)). Long-term stability measurements under repeated on/off illumination cycles over 15 min are shown in Fig. S7, demonstrating stable and reproducible photoresponse throughout the measurement period. The rise and fall times were defined as the time required for the photocurrent to change between 10% and 90% of its maximum value. These measurements showed that the device fabricated on m-plane sapphire exhibited a substantially slower response, with rise and fall times of 14.70 and 3.55 s, respectively, compared with 0.40/0.18 s for the r-plane and 0.51/0.14 s for the a-plane devices. The approximately twofold difference in photocurrent between Fig. 6(f) and (i) likely arises from the slow response of the m-plane device, which may not fully reach steady state during the 20 s illumination interval, whereas the I–V measurement was performed after prolonged illumination. The noticeably slower photocurrent stabilization of the m-plane device, together with its sublinear power dependence ($\gamma = 0.71$) and persistent photoconductivity component observed in Fig. S7(c), is consistent with a significant contribution from carrier trapping and detrapping processes. Importantly, the high responsivity of the m-plane device is therefore accompanied by a pronounced responsivity–speed trade-off. In photoconductors, trapping of one carrier type can prolong the lifetime of the opposite carrier, allowing repeated carrier transport through the external circuit before recombination and thereby producing photoconductive gain; the same long-lived trapped carriers can, however, substantially increase the response and recovery times [42,43]. Such a defect-mediated gain mechanism provides a plausible explanation for the simultaneous ~three-orders-of-magnitude enhancement in responsivity and slower temporal response observed for the m-plane device, with the EQE > 100% further supporting the presence of photoconductive gain. Interestingly, this film also exhibited the distinctive and intense PL component at ~3.15 eV discussed above, suggesting that the defect states associated with this emission may contribute to the trapping processes responsible for the enhanced photoconductive gain. However, establishing a direct correspondence between this PL feature and the electrically active trap states would require a dedicated defect spectroscopy study. Analogous correlations between specific defect-related PL emission and trap-mediated photoconductive gain have been previously reported for amorphous gallium oxide photodetectors [43].

The peak responsivity of 1.60 A/W achieved by the m-plane r-$GeO_2$ device is comparable in magnitude to values reported for high-performance UWBG photoconductive detectors based on materials such as AlGaN, β-$Ga_2O_3$, and diamond [3,44–46]. However, its slower temporal response and the still poorly understood role of structural and point defects highlight the need for further studies aimed at reducing extended defects and controlling electrically active point defects through growth optimization and post-deposition processing. At this stage, the main significance of r-$GeO_2$ lies in its emerging photoconductive physics and intrinsic UV-C spectral selectivity rather than in direct performance competition with more mature UWBG detector platforms.

**Table 1.** Summary of the photodetection performance of as-fabricated r-$GeO_2$ thin-film MSM devices on different sapphire substrate orientations at 5 V bias.

| Rutile $GeO_2$ thin film | Peak $R_\lambda$, mA/W | EQE, % | D*, Jones | Solar-blind RR, $R_{peak}/R_{280nm}$ | Visible-blind RR, $R_{peak}/R_{400nm}$ | Power-law exponent γ | $I_{dark}$, pA | $I_{photo}/I_{dark}$ | $t_{rise}$, s | $t_{fall}$, s |
|---|---|---|---|---|---|---|---|---|---|---|
| On r-plane sapphire | 5.8 at 225 nm | 3 | $4.4 \cdot 10^{10}$ | 9.2 | 41.2 | 0.75 | 38 | 50 | 0.40 | 0.18 |
| On a-plane sapphire | 1.8 at 230 nm | 1 | $7.4 \cdot 10^{10}$ | 4.5 | 37.5 | 0.79 | 10 | 95 | 0.51 | 0.14 |
| On m-plane sapphire | 1603.7 at 220 nm | 902 | $4.6 \cdot 10^{12}$ | 34.1 | 153.4 | 0.71 | 750 | 413 | 14.70 | 3.55 |

## Conclusions

We systematically investigated the influence of sapphire substrate orientation on the PLD growth, material properties, and solar-blind photodetection performance of phase-pure r-$GeO_2$ thin films. The r- and a-plane sapphire resulted in preferentially (101)-oriented films, whereas the m-plane sapphire promoted (002)-oriented growth with the lowest rocking-curve FWHM of 0.46°, a comparatively well-ordered film–substrate interface, and the lowest surface roughness of 2.70 nm. These results demonstrate the critical role of substrate orientation in controlling the crystalline quality and morphology of PLD-grown r-$GeO_2$. The corresponding m-plane photodetector exhibited a peak responsivity of 1.60 A/W and a solar-blind rejection ratio (relative to the responsivity at 280 nm) of 34.1, substantially outperforming the devices fabricated on r- and a-plane sapphire. The enhanced responsivity was accompanied by considerably slower temporal response, indicating a pronounced responsivity–speed trade-off consistent with defect-mediated photoconductive gain. The distinctive ~3.15 eV PL emission observed for this film suggests a possible connection with the relevant trap states, although dedicated defect spectroscopy is required to establish their microscopic origin. Optical and photoconductive characterization revealed an effective absorption and photoresponse onset at photon energies considerably higher than the fundamental electronic bandgap of ~4.7–4.8 eV. The experimentally obtained optical absorption edge of ~5.1 eV, together with the blue-shifted photoconductive onset, is consistent with theoretical predictions of an optically weak or dipole-forbidden fundamental transition in r-$GeO_2$. Suppression of photogeneration close to the fundamental band edge effectively extends the intrinsic spectral selectivity deeper into the UV-C region, thereby providing an important advantage of r-$GeO_2$ for solar-blind photodetection. Overall, these results establish substrate orientation as a key parameter for tailoring PLD-grown r-$GeO_2$ and highlight the material's potential for next-generation solar-blind UV-C optoelectronics.

**Supporting Information:** Optical microscope image of the photodetector electrode geometry; XRD patterns of bare sapphire substrates; growth-temperature-dependent SEM and AFM analysis; detailed HR-TEM and SAED crystallographic characterization of the films and film–substrate interfaces; comparative XPS Ge 3d and O 1s spectra; UV–Vis–NIR absorptance, transmittance, and reflectance spectra; and extended photocurrent on/off cycling measurements.

**Funding:** This research was funded by the European Regional Development Fund Project No. 1.1.1.3/1./24/A/020. E.D. acknowledges financial support by the Project No. 1.1.1.8/1/24/I/003 “Strengthening the Research and Development Capacity of Doctoral Studies at the University of Latvia in the Fields of Smart Specialization”. E.I. acknowledges financial support provided by the "Post-doctoral Research" project No. 1.1.1.9/LZP/2/25/197.

**Acknowledgments:** Generative artificial intelligence was used solely for language editing and proofreading of the manuscript. The authors reviewed and approved all resulting changes and take full responsibility for the final content.

**Author Contributions:** Conceptualization, E.B.; Methodology, E.D. and E.B.; Validation, E.D., E.I., E.B.; Investigation, E.D., E.S., S.O., A.T., K.L., E.V., J.G., B.P., E.I. and E.B.; Writing – Original Draft Preparation, E.D. and E.B.; Writing – Review & Editing, E.D., E.S., S.O., A.T., K.L., E.V., B.P. and E.B.; Visualization, E.D. and E.B.; Supervision, E.B.; Project administration, E.D., E.I. and E.B.

**Data Availability Statement:** The data supporting this study's findings are available from the corresponding author upon reasonable request.

**Conflicts of Interest:** The authors declare no competing financial interest.